\documentclass[preprint,12pt]{elsarticle}

\input epsf
\usepackage{amssymb}

\newcounter{bla}

\def\tfrac#1#2{{{\lower.6ex
\hbox{$\scriptstyle#1$}}\over
{\raise.7ex
\hbox{$\scriptstyle#2$}}}}

\def\phase{{\rm ph}}

\def\dsp#1{\displaystyle#1}

\def\sk{\sum_{k=0}^\infty\,}

 \def\protectbold#1{\protect{\boldmath{$#1$}}}

\def\Frac#1#2{\frac{\displaystyle{#1}}{\displaystyle{#2}}}

\journal{Computer Physics Communications}

\begin{document}

\begin{frontmatter}



\title{{\bf Conical}: an extended module for computing a numerically satisfactory pair of
solutions of the differential equation for conical functions}

\author[1]{T.M. Dunster}
\author[2,1]{A. Gil}
\author[3,1]{J. Segura}
\author[4]{N.M. Temme}
\address[1]{Department of Mathematics and Statistics. San Diego State University. 5500 Campanile Drive San Diego, CA, USA.}
\address[2]{Depto. de Matem\'atica Aplicada y Ciencias de la Comput. Universidad de Cantabria. 39005-Santander, Spain}
\address[3]{Depto. de Matem\'aticas, Estad\'{\i}stica y Comput. Universidad de Cantabria. 39005-Santander, Spain}
\address[4]{ IAA, 1825 BD 18, Alkmaar, The Netherlands\footnote{Former address: CWI, 1098 XG Amsterdam, The Netherlands} }
\begin{abstract}

Conical functions appear in a large 
number of applications in physics and engineering.
In this paper we describe an extension of our module {\bf Conical}  \cite{gil:2012:cpc}  for the
computation of conical
functions. Specifically, the module includes now a routine for computing 
the function  ${{\rm R}}^{m}_{-\frac{1}{2}+i\tau}(x)$, a real-valued numerically
satisfactory companion of the function ${\rm P}^m_{-\tfrac12+i\tau}(x)$ for $x>1$.
In this way, a natural basis for solving Dirichlet problems bounded by conical
domains is provided.

The module also improves the performance of our previous algorithm for the conical function ${\rm P}^m_{-\tfrac12+i\tau}(x)$
and it includes now the computation of the first order derivative of the function. This is also considered for the function
${{\rm R}}^{m}_{-\frac{1}{2}+i\tau}(x)$ in the extended algorithm.
\end{abstract}
\end{frontmatter}

\section{Introduction}

Conical or Mehler functions are involved in a large number of applications in different areas of physics. 
In particular, these functions appear when solving the Laplace
equation in spherical coordinates for two intersecting cones \cite{thebault:2006:geo}
 or for regions bounded by two intersecting spheres, or by one or two confocal hyperboloids of revolution 
when using toroidal coordinates.  

An extended version of our module {\bf Conical}  \cite{gil:2012:cpc}  for the
computation of conical
functions is presented in this paper. The new module includes now a routine for computing 
the function  ${{\rm R}}^{m}_{-\frac{1}{2}+i\tau}(x)$, a real-valued numerically
satisfactory companion of the function ${\rm P}^m_{-\tfrac12+i\tau}(x)$ for $x>1$.
The module also improves our previous algorithm for the conical function ${\rm P}^m_{-\tfrac12+i\tau}(x)$
by considering now more cofficients in some of the asymptotic expansions used for computing the function  in the region $x>1$.
 The computation of the first order derivatives of ${\rm P}^m_{-\tfrac12+i\tau}(x)$ and  ${{\rm R}}^{m}_{-\frac{1}{2}+i\tau}(x)$ 
is also included in the new module. 

\section{Theoretical background}

  Conical functions  are solutions of the second order differential equation

\begin{equation}\label{ODE}
(1-x^2) \Frac{d^2 w}{dx^2}-2x \Frac{dw}{dx}-\left(\tau^2+\Frac{1}{4}+\Frac{\mu^2}{1-x^2}  \right)w=0.
\end{equation}
   
We will restrict to integer positive values of the parameter $\mu$ ($\mu=m \in {\mathbb Z}^+$).

  When $-1<x<1$, a real-valued satisfactory pair of solutions of (\ref{ODE}) is
${\rm P}^{m}_{-\frac{1}{2}+i\tau}(x)$ and ${\rm P}^{-m}_{-\frac{1}{2}+i\tau}(x)$.
Both functions can be computed using our algorithm for  ${\rm P}^{m}_{-\frac{1}{2}+i\tau}(x)$ implemented in {\bf conicp};
for computing ${\rm P}^{-m}_{-\frac{1}{2}+i\tau}(x)$ the following relation can be used

\begin{equation}
\label{changem}
{\rm{P}}^{-m}_{-\frac12+i\tau}(x)=\Frac{\pi}{\cosh(\pi\tau)\left|\Gamma (m+\tfrac12+i\tau)\right|^2}
{\rm{P}}^{m}_{-\frac12+i\tau}(x).
\end{equation}

 When $x>1$, a real-valued satisfactory pair of solutions of (\ref{ODE}) is
${\rm P}^{m}_{-\frac{1}{2}+i\tau}(x)$ and ${\rm R}^{m}_{-\frac{1}{2}+i\tau}(x) 
\equiv \Re\left\{e^{-i\pi m} {{\rm Q}}^{m}_{-\frac{1}{2}+i\tau}(x)    \right\}$
(the function ${{\rm Q}}^{m}_{-\frac{1}{2}+i\tau}(x)$ is complex-valued). 

The Wronskian relation for ${\rm P}^{m}_{-\frac{1}{2}+i\tau}(x)$ and ${\rm R}^{m}_{-\frac{1}{2}+i\tau}(x)$, useful for testing,   is given by

\begin{equation}
\label{wronski}
{{W}\/}\left\{{{\rm P}^{m}_{-\frac{1}{2}+%
i\tau}\/}\!\left(x\right),{{{\rm R}}^{m}_{-\frac%
{1}{2}+i\tau}\/}\!\left(x\right)\right\}=\frac{\pi(e^{-\tau\pi}
+{\sinh\/}\!%
\left(\tau\pi\right))}{|{\Gamma\/}\!\left(-m+\frac{1}{2}+i%
\tau\right)|^{2}({{\cosh\/}^{2}}\!\left(\tau\pi\right))(1-x^{2})}\,.
\end{equation}

The algorithm for computing the conical function ${\rm P}^{m}_{-\frac{1}{2}+i\tau}(x)$  was described in
\cite{gil:2012:cpc}.
In the new module {\bf Conical} we also compute the first order derivatives for  ${\rm P}^m_{-\tfrac12+i\tau}(x)$ and 
${{\rm R}}^{m}_{-\frac{1}{2}+i\tau}(x)$: the first order derivative of ${\rm P}^m_{-\tfrac12+i\tau}(x)$  can be obtained  
using the relation

\begin{equation}
\label{deriv}
\Frac{d}{dx}{\rm P}^m_{-\tfrac12+i\tau}(x)=
-\Frac{1}{\sqrt{x^2-1}}{\rm P}^{m+1}_{-\tfrac12+i\tau}(x)+
\Frac{mx}{x^2-1}{\rm P}^{m}_{-\tfrac12+i\tau}(x)\,.
\end{equation}

The first order derivative of ${{\rm R}}^{m}_{-\frac{1}{2}+i\tau}(x)$ satisfies the same relation (\ref{deriv})
with ${\rm R}^{m+1}_{-\tfrac12+i\tau}(x)$ and ${\rm R}^{m}_{-\tfrac12+i\tau}(x)$.

 A preliminary algorithm for computing the function  ${\rm R}^{m}_{-\frac{1}{2}+i\tau}(x)$
was presented in \cite{dunster:2015:con} although the final algorithm in finite precision arithmetic
implemented in the routine {\bf conicr} presents some differences with respect to the first algorithm.  
We have also changed the notation of the function with respect to the one used in that reference; we are using 
now ${\rm R}^{m}_{-\frac{1}{2}+i\tau}(x) $ instead of $\widetilde{Q}^{m}_{-\frac{1}{2}+i\tau}(x)$ for simplicity.  

 Next we summarize the methods for computing the conical function ${\rm R}^{m}_{-\frac{1}{2}+i\tau}(x)$.

    \subsection{Computation of   ${\rm R}^{m}_{-\frac{1}{2}+i\tau}(x)$  for values of $x$ close to $1$}

\subsubsection{Small or moderate values of $\tau$}
\label{r0r1}
   To compute ${\rm R}^{0}_{-\frac{1}{2}+i\tau}(x)$  and  ${\rm R}^{1}_{-\frac{1}{2}+i\tau}(x)$   we will use the expansions

\begin{equation}
\label{eq:Q01}
\begin{array}{lcl}
{\rm R}_{-\frac12+i\tau }^{\,0}(x)&=&
\Re \left( \dsp{\sum_{k=0}^{\infty} \frac{\left(\tfrac12-i\tau\right)_k\left(\tfrac12+i\tau\right)_k }{(k!)^2}z^k\bigl(\psi(k+1)-\psi(\tfrac12+i\tau)-\ln w\bigr)}\right),\\
{\rm R}_{-\frac12+i\tau }^{\,1}(x)&=&\Re \left(   \sqrt{x^2-1} 
 \dsp{\sum_{k=0}^{\infty}
 \frac{\left(\tfrac12-i\tau\right)_k\left(\tfrac12+i\tau\right)_k }{(k!)^2}\tfrac12 z^{k-1}\right.
\ \left(   \tfrac12  \left(-1+w^2   \right)  }  + \right. \\
&&\left. \left. k \bigl(\psi(k+1)-\psi(\tfrac12+i\tau)-\ln w\bigr)  \right) \right),
\end{array}
\end{equation}
where $z$ and $w$ are given by

\begin{equation}
\label{zeta}
z=\tfrac12(1-x), \,\,w=\dsp{\sqrt{\frac{x-1}{x+1}}},
   \end{equation}
and $\psi(\alpha)=\Gamma^\prime(\alpha)/\Gamma(\alpha)$.  

The only complex quantity in (\ref{eq:Q01}) is the function $\psi(\tfrac12+i\tau)$. For computing this function we use an algorithm which computes separately
the real and imaginary parts of  $\psi(\tfrac12+i\tau)$ avoiding, in this way, the use of complex arithmetics when computing (\ref{eq:Q01}). The algorithm
is based on the use of the asymptotic expansion
\begin{equation}\label{eq:QKumF08}
\psi(\alpha)\sim\dsp{\ln \alpha}-\frac{1}{2\alpha}-\frac{1}{12\alpha^2}+\frac{1}{120\alpha^4}-\frac{1}{252\alpha^6}+...,
\end{equation}
valid for $\alpha\to\infty$ in $\vert\phase\,\alpha\vert<\pi$. We use this expansion if $\vert \alpha\vert \ge 12$ with $8$ terms 
of the series (or less), and we use the recurrencence relation $\psi(\alpha)=\psi(\alpha+1)-1/\alpha$ 
for smaller values of $\vert \alpha\vert$. 

When $m \ge 2$ we will use the recursion relation 

\begin{equation}
\label{TTRR}
{\rm R}_{-\frac12+i\tau}^{m +1}(x)-\Frac{2m x}{\sqrt{x^2-1}}
{\rm R}_{-\frac12+i\tau}^{m}(x) +
\left((m-\tfrac12)^2+\tau^2\right){\rm R}_{-\frac12+i\tau}^{m-1}(x)  =0
\end{equation}
in the direction of increasing $m$.   

\subsubsection{Large values of $\tau$}
\label{kummer}
A representation in terms of Kummer   \protectbold{U-}functions is used in this case:

\begin{equation}\label{eq:QKum15}
\begin{array}{@{}r@{\;}c@{\;}l@{}}
{\rm R}_{-\frac12+i\tau}^\mu(x)
&\sim&
\dsp{\sqrt{\pi/2}\, \alpha^{\mu+\frac12}
\left(x^2-1\right)^{-\frac14}\ \times}\\[8pt]
&&\dsp{ \sk f_k   \left(    \Re \Phi_k \cos \phi + \Im   \Phi_k  \sin \phi     \right)                  ,}
\end{array}
\end{equation}
where $\phi=\tau\log(x+\sqrt{x^2-1})$,  $\alpha=\ln \left(  \Frac{z+1}{z}   \right)$  and $z$ is given by

\begin{equation}\label{eq:QKum02}
z=\dsp{\frac{1}{2\sqrt{x^2-1}\left(x+\sqrt{x^2-1}\right)}.}
\end{equation}

The functions $\Phi_k$ are given in terms of Kummer  \protectbold{U-}functions as follows

\begin{equation}\label{eq:QKum16}
\Phi_k= \left(\tfrac12-\mu\right)_k \omega^{2\mu-k}U\left(\tfrac12+\mu,1+2\mu-k,\alpha \omega\right),
\quad \omega=i\tau.
\end{equation}

The functions can be also written in terms of the Hankel functions  $H_\mu^{(2)}(z)$. This representation makes simple separating
the real and imaginary parts of $\Phi_k$ by using

\begin{equation}\label{eq:QKum19}
H_\mu^{(2)}(z)=J_\mu(z)-iY_\mu(z).
\end{equation}

For the computation of the Bessel functions $J_\mu(z)$, $Y_\mu(z)$ we use an algorithm which combines
the use of  series expansions, Debye's asymptotic expansions, asymptotic expansions for large $z$, Airy-type
asymptotic expansions and three-term recurrence relations. This algorithm is implemented in the module {\bf BesselJY}
and it is also included in the software package.

The functions $\Phi_0$, $\Phi_1$ are given by

\begin{equation}\label{eq:QKum17}
\begin{array}{lcl}
\Phi_0&=&-\tfrac12i\sqrt{\pi}(\tau/\alpha)^\mu e^{\frac12i\alpha\tau}H_\mu^{(2)}\left(\tfrac12\alpha\tau\right),\\
\Phi_1&=& \dsp{\tfrac14\alpha\sqrt{\pi}(\tau/\alpha)^\mu e^{\frac12i\alpha\tau}} \left(iH_\mu^{(2)}\left(\tfrac12\alpha\tau\right)
+H_{\mu-1}^{(2)}\left(\tfrac12\alpha\tau\right)\right).
\end{array}
\end{equation}

For computing $\Phi_n$ for $n=2,...$ we can use a recurrence relation for the Kummer \protectbold{U-}functions which gives

\begin{equation}\label{eq:QKum10}
\omega\Phi_{n+1}=(n-2\mu-\alpha\omega)\Phi_n+\alpha(n-\tfrac12 -\mu)\Phi_{n-1}.
\end{equation}

From this, the following recurrence relations for both the real and  imaginary parts of $\Phi_{n+1}$ can be obtained:

\begin{equation}\label{eq:QKum11}
\begin{array}{lcl}
\Re\Phi_{n+1}&=&\Frac{n-2\mu}{\tau}\Im \Phi_n-\alpha \Re \Phi_n+\Frac{\alpha}{\tau}(n-\tfrac12 -\mu)\Im \Phi_{n-1},\\
\Im\Phi_{n+1}&=&-\Frac{n-2\mu}{\tau}\Re \Phi_n-\alpha \Im \Phi_n-\Frac{\alpha}{\tau}(n-\tfrac12 -\mu)\Re \Phi_{n-1}.
\end{array}
\end{equation}

The first coefficients $f_k$ in (\ref{eq:QKum15}) are given by

\begin{equation}
\begin{array}{lcl}
f_0&=&1,\\[6pt]
f_1&=&\Frac{b}{2d}\left(2dz+d-2z\right),\\[6pt]
f_2&=&\Frac{b}{24d^2}\left(12z^2+12bz^2+d^2-12d^2z-12d^2z^2-24bdz^2\right.\\[6pt]
     &&\left.\,\,\,\,\,\,\,\,\,\, +12bd^2z+12bd^2z^2+3bd^2-12bdz\right),
\end{array}
\end{equation}
where $b=-\mu-\frac12$ and $d=z \alpha$.

 \subsection{Computation of   ${\rm R}^{m}_{-\frac{1}{2}+i\tau}(x)$  for moderate or large values of $x$}
\label{largex}

In this case we use the expansion

\begin{equation}\label{eq:qhyp1}
\begin{array}{@{}r@{\;}c@{\;}l@{}}
  {\rm R}_{-\frac12+i\tau}^\mu(x)  &=& \Re \left(
 \dsp{\sqrt{\pi/2} \,\left(x^2-1\right)^{-\frac14}G(\mu,\tau)\,e^{-i\phi}} \right.
\ \times \\
&&\left. \dsp{\sum_{k=0}^{\infty} \left(\tfrac12+\mu \right)_k\left(\tfrac12-\mu \right)_k 
\frac{u_k(\tau)+iv_k(\tau)}{w_k(\tau)}\Frac{(-z)^k}{k!} \right),                                           }
\end{array}
\end{equation}
where $z$ is given in eq.(\ref{eq:QKum02}),  $\phi=\tau\log\left(x+\sqrt{x^2-1}\right)$, 

\begin{equation}
\label{gmu}
G(\mu,\tau)=\Frac{\Gamma\left(\tfrac12+\mu+i\tau\right)}{\Gamma\left(1+i\tau\right)}, 
\end{equation}
and
\begin{equation}
\label{eq:qhyp2}
\frac{u_k(\tau)+iv_k(\tau)}{w_k(\tau)}=\frac{1}{\left(1+i\tau\right)_k}, \quad k=0,1,2,\ldots.
\end{equation}
We can compute $u_k(\tau)$, $v_k(\tau)$ and $w_k(\tau)$ from the recurrence relations
\begin{equation}
\label{eq:qhyp3}
\begin{array}{@{}r@{\;}c@{\;}l@{}}
u_{k+1}(\tau)&=&(k+1)u_k(\tau)+\tau v_k(\tau),\\[8pt]
v_{k+1}(\tau)&=&(k+1)v_k(\tau)-\tau u_k(\tau),\\[8pt]
w_{k+1}(\tau)&=&\left((k+1)^2+\tau^2\right)w_k(\tau),
\end{array}
\end{equation}
with $u_0(\tau)=1$, $v_0(\tau)=0$, $w_0(\tau)=1$.

The real part of (\ref{eq:qhyp1}) can be obtain by writing

\begin{equation}
\label{eq:qhyp4}
G(\mu,\tau)=H(\mu,\tau)e^{i\rho(\mu,\tau)},\quad u_k(\tau)+iv_k(\tau)=r_k(\tau)e^{i\sigma_k(\tau)},
\end{equation}
which gives
\begin{equation}
\label{eq:qhyp5}
\begin{array}{lcl}
{\rm R}_{-\frac12+i\tau}^\mu(x)  &=&
 \dsp{\sqrt{\pi/2}\, H(\mu,\tau)\,\left(x^2-1\right)^{-\frac14}\,
\ \times} \\
&&\dsp{\sum_{k=0}^{\infty} \left(\tfrac12+\mu \right)_k\left(\tfrac12-\mu \right)_k 
\Frac{r_k(\tau)}{w_k(\tau)}\Frac{(-z)^k}{k!}\cos(\psi_k}),                                           
\end{array}
\end{equation}
where

\begin{equation}
\label{eq:qhyp6}
\psi_k =\tau\log\left(x+\sqrt{x^2-1}\right)-\rho(\mu,\tau)-\sigma_k(\tau).
\end{equation}

The computation of the ratio of two gamma functions in (\ref{gmu}) is made using an algorithm
for computing the gamma function for complex values of the argument. The algorithm adapts for complex arguments
the scheme used for real values described in \cite{gil:2015:CHI}.

\section{Overview of the software structure}

The Fortran 90 package includes the main module {\bf Conical}, which includes
the routines {\bf conicp}, {\bf conicr} and {\bf conicpr}.

In the module {\bf Conical}, the auxiliary 
module {\bf Someconstants} is used. This is a module for the 
computation of the main constants used in 
  the different routines. 
The module {\bf BesselJY} (for the computation of Bessel functions) and {\bf AiryFunction} 
(for the computation of Airy functions) are used. 
The routines included in {\bf auxil.f90} are also used in the module
{\bf Conical}.

\section{Description of the individual software components}

The Fortran 90 module {\bf Conical} includes the public routine
{\bf conicp}, which computes the conical functions ${\rm P}^m_{-\frac12+i\tau}(x)$ for $x>-1$,
 $m\ge 0$ and $\tau >0$;
the routine {\bf conicr},  which computes
the function  ${{\rm R}}^{m}_{-\frac{1}{2}+i\tau}(x)$, for $x>1$ , $m\ge 0$ and $\tau >0$
and the routine  {\bf conicpr}, which computes both functions ${\rm P}^m_{-\frac12+i\tau}(x)$, 
 ${{\rm R}}^{m}_{-\frac{1}{2}+i\tau}(x)$ and their first order derivatives for $x>1$ , $m\ge 0$ and $\tau >0$.
The calling sequences of these routines are

  \begin{verbatim}
   CALL  conicp(x,mu,tau,pm,ierr)
  \end{verbatim}

   \noindent
   where the input data are: $x$, $mu$ and $tau$ (arguments of the functions). 
The outputs are the
error flag $ierr$ and the function value $pm$.
The possible values of the error flag are: $ierr=0$, successful 
computation; 
$ierr =1$, computation failed due to overflow/underflow; 
$ierr=2$, arguments out of range.

  \begin{verbatim}
   CALL  conicr(x,mu,tau,rm,ierr)
  \end{verbatim}

   \noindent
   where the input data are: $x$, $mu$ and $tau$ (arguments of the functions).
The outputs are the
error flag $ierr$ and the function value $rm$.
The possible values of the error flag are: $ierr=0$, successful 
computation; 
$ierr =1$, computation failed due to overflow/underflow; 
$ierr=2$, arguments out of range.  

  \begin{verbatim}
   CALL  conicpr(x,mu,tau,pm,pmd,rm,rmd,ierr)
  \end{verbatim}

   \noindent
   where the input data are: $x$, $mu$ and $tau$ (arguments of the functions).
The outputs are the
error flag $ierr$, the function values $pm$, $rm$ and the first order derivatives $pmd$, $rmd$.
The possible values of the error flag are: $ierr=0$, successful 
computation; 
$ierr =1$, computation failed.

\section{Testing the algorithm}

 For testing the accuracy of the expansions used to compute the 
conical function  ${\rm R}^{m}_{-\frac{1}{2}+i\tau}(x)$, we have first checked
 (\ref{TTRR}) written in the form

\begin{equation}
\Frac{\left(2mx/\sqrt{x^2-1}\right){{\rm R}}^{m}_{-\frac{1}{2}+i\tau}(x)-\left((m-\tfrac12)^2+\tau^2\right){\rm R}_{-\frac12+i\tau}^{m-1}(x)}
{{{\rm R}}^{m+1}_{-\frac{1}{2}+i\tau}(x) }=1.      
\label{errRR1}
\end{equation}

This check fails close to the zeros of ${\rm R}^{m+1}_{-\frac{1}{2}+i\tau}(x)$; in this case, we can consider the
alternative test

\begin{equation}
\Frac{{{\rm R}}^{m+1}_{-\frac{1}{2}+i\tau}(x)+\left((m-\tfrac12)^2+\tau^2\right){\rm R}_{-\frac12+i\tau}^{m-1}(x)}
{\left(2mx/\sqrt{x^2-1}\right){{\rm R}}^{m}_{-\frac{1}{2}+i\tau}(x)}=1.      
\label{errRR2}
\end{equation}

\begin{figure}
\begin{center}
\hspace*{-1.7cm}
\epsfxsize=17cm \epsfbox{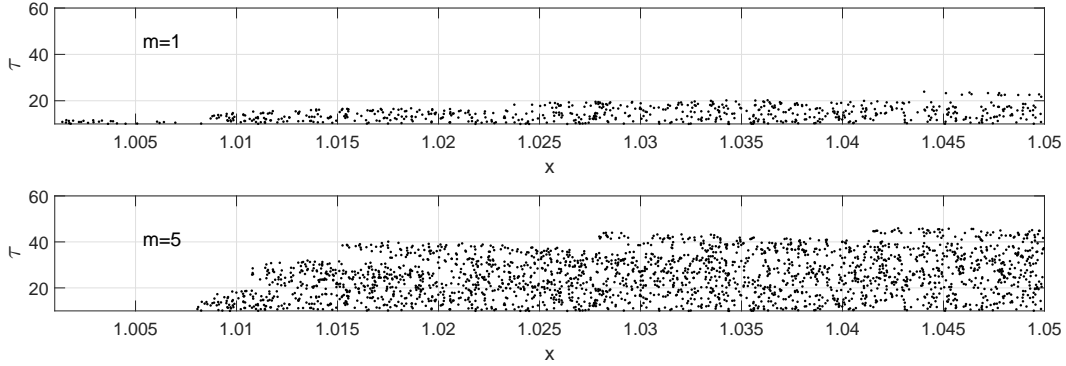}
\caption{Test of the performance of the expansion (\ref{eq:QKum15}). The points where the value of the error
when testing the recurrence relation (\ref{TTRR}) is greater than
$10^{-12}$ are plotted. 
\label{Fig1}}
\end{center}
\end{figure}

Because the zeros of ${\rm R}^{m+1}_{-\frac{1}{2}+i\tau}(x)$ and ${\rm R}^{m}_{-\frac{1}{2}+i\tau}(x)$
are interlaced, both tests will not fail simultaneously. We can therefore take the minimum of both errors.
We have considered these tests for the expansions described in sections \ref{kummer} ($x$ small)
and \ref{largex} ($x$ large).
Figure \ref{Fig1} shows the points where the minimum value of the error
of the tests (\ref{errRR1}) and  (\ref{errRR2}) when using  (\ref{eq:QKum15}) is greater than $10^{-12}$. In the 
algorithm, we have fixed to $N=7$ the number of terms used in the expansion. Random points have been generated
in the domain $(x,\,\tau) \in (1.001,\,1.05) \times (15,\,100)$. As can be seen, for $m=1$ (upper figure) the use of the expansion (\ref{eq:QKum15}) allows to compute the function values with an accuracy better than $10^{-12}$ for values of $\tau$ greater than $20$ when $x$ is close to $1$. The accuracy of the expansion worsens as $m$ increases, as can be seen also in Figure \ref{Fig1} (lower figure) where the same test is considered for $m=5$.  Therefore, one has to use an alternative method of computation for
moderate/large values of $m$ as, for example, the use of the recurrence relation  (\ref{TTRR}) starting from ${\rm R}^{0}_{-\frac{1}{2}+i\tau}(x)$ and  ${\rm R}^{1}_{-\frac{1}{2}+i\tau}(x)$. 

\begin{figure}
\begin{center}
\hspace*{-1.6cm}
\epsfxsize=17cm \epsfbox{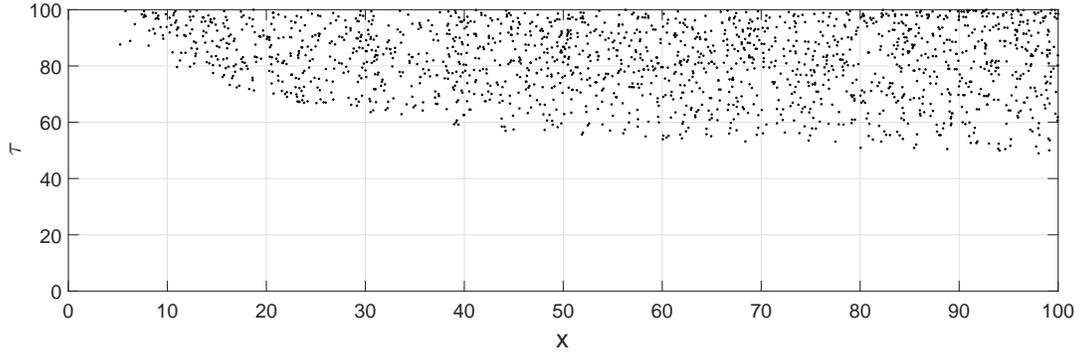}
\caption{Test of the performance of the expansion (\ref{eq:qhyp1}) for $\mu \equiv m=1$. The points where the value of the error in the recurrence relation  (\ref{TTRR}) is greater than
$10^{-12}$ are plotted.
\label{Fig2} }
\end{center}
\end{figure}

Figure \ref{Fig2} shows the same tests  (\ref{errRR1}) and  (\ref{errRR2})  for the expansion (\ref{eq:qhyp1}) and for $\mu \equiv m=1$. The domain where the random points have been generated is now $(x,\,\tau) \in (1.2,\,100) \times (0,\,100)$. As can be seen in the figure, there is some loss of accuracy when $\tau$ is large and $x$ is moderate/large. In any case, we have checked that the accuracy
was always better than $5\,10^{-12}$.

For testing the expansions for  ${\rm R}^{0}_{-\frac{1}{2}+i\tau}(x)$ and  ${\rm R}^{1}_{-\frac{1}{2}+i\tau}(x)$ of
section \ref{r0r1}, we have used the Wronskian relation given in (\ref{wronski}). In this case, we have

\begin{equation}
\label{wronski2}
{\rm P}^{1}_{-\frac{1}{2}+i\tau}(x){\rm R}^{0}_{-\frac{1}{2}+i\tau}(x)-{\rm P}^{0}_{-\frac{1}{2}+i\tau}(x){\rm R}^{1}_{-\frac{1}{2}+i\tau}(x)=
\Frac{e^{-\pi \tau} +\sinh (\pi \tau)}{\cosh (\pi \tau)\sqrt{x^2-1}}\,,
\end{equation}

\begin{figure}
\begin{center}
\hspace*{-1.6cm}
\epsfxsize=17cm \epsfbox{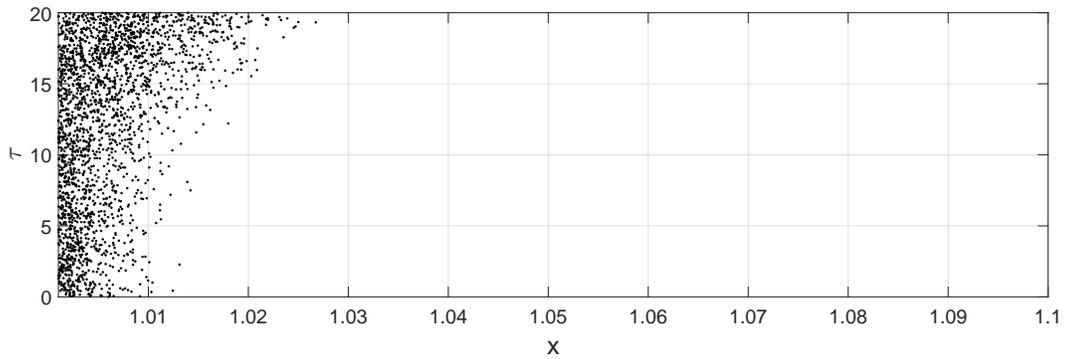}
\caption{Test of the performance of the expansions given in (\ref{eq:Q01}). The points where the value of the error in the Wronskian relation
(\ref{wronski2}) is greater than
$10^{-12}$ are plotted.
\label{Fig3} }
\end{center}
\end{figure}

Figure \ref{Fig3} shows the points where the value of the error in the Wronskian relation
(\ref{wronski2}) is greater than $10^{-12}$.

  The accuracy of the final algorithm for  ${\rm R}^{m}_{-\frac{1}{2}+i\tau}(x)$ has been tested by computing the Wronskian relation given
in (\ref{wronski}) for a very large number of random points in the parameter domain $(x,\,m\,\tau) \in (1.001,\,100) \times  (0,\,100) \times  (0,\,100)$.
The algorithm for the conical function ${\rm P}^m_{-\tfrac12+i\tau}(x)$ was improved
by considering more cofficients in some of the asymptotic expansions used for computing the function  in the region $x>1$.
We have checked that the accuracy of the Wronskian test (\ref{wronski})  is close to $10^{-12}$ in the whole parameter domain 
and better than $10^{-13}$ for a large fraction of the 
tested parameter values.

\section{Test run description}

The Fortran 90 test program {\bf testcon.f90} includes the computation
of 20 values of the functions ${\rm P}^m_{-\tfrac12+i\tau}(x)$,  ${\rm R}^{m}_{-\frac{1}{2}+i\tau}(x)$ and their first order
derivatives and their comparison with the corresponding
pre-computed results.

\section{Acknowledgements}

A.G. acknowledges the Fulbright/MEC Program for support during her stay at SDSU.
J.S.  acknowledges the Salvador de Madariaga Program for support during his stay at SDSU.
The authors acknowledge financial support from 
{\emph{Ministerio de Ciencia e Innovaci\'on}}, project MTM2015-67142-P. NMT thanks CWI, Amsterdam, for scientific support.





\section*{References}

\bibliographystyle{elsarticle-num}
\bibliography{ConicalCPC}



\end{document}